\RequirePackage{lineno}
\documentclass[aps,prd,twocolumn,showpacs,preprintnumbers,amsmath,amssymb,superscriptaddress]{revtex4-1}
 \usepackage{graphicx}
 \usepackage{amsmath}
 \usepackage{amssymb}
 \usepackage{amstext}
 \usepackage{nicefrac}
 \usepackage{afterpage}
 \usepackage{graphics}
 \usepackage{float}
 \usepackage{rotating}
 \usepackage{xspace}
 \usepackage{dcolumn}%align table columns on decimal point
 \usepackage{bm} %bold math
 \usepackage{color}
 \usepackage{appendix}
 
\newcommand{\nova}{NOvA\xspace}

\newcommand{\realpot}{\ensuremath{3.45\,\mathord{\times}\,10^{20}} protons-on-target\xspace}
\newcommand{\effpot}{\ensuremath{2.74\,\mathord{\times}\,10^{20}} protons-on-target\xspace}
\newcommand{\effpotPOT}{\ensuremath{2.74\,\mathord{\times}\,10^{20}} POT\xspace}
\newcommand{\ndpot}{\ensuremath{1.66\,\mathord{\times}\,10^{20}}\xspace}
\newcommand{\mus}{$\mu$s\xspace}

\newcommand{\Ehad}{\ensuremath{E_{\mathrm{had}}}\xspace}

\newcommand{\numu}{\ensuremath{\nu_{\mu}}\xspace}                   % nu_mu
\newcommand{\nue}{\ensuremath{\nu_{e}}\xspace}                      % nu_e
\newcommand{\nutau}{\ensuremath{\nu_{\tau}}\xspace}                 % nu_tau
\begin{document}
\setpagewiselinenumbers
\modulolinenumbers[1]
%\linenumbers


%\leftline{Version Draft.1 as of \today}
%\leftline{To be submitted to PRD RC}

%\title{First results on muon neutrino disappearance from the \nova experiment}
\title{First measurement of muon-neutrino disappearance in \nova}

\newcommand{\ANL}{Argonne National Laboratory, Argonne, Illinois 60439, 
USA}
\newcommand{\IOP}{Institute of Physics, The Czech Academy of Sciences, 
Prague, Czech Republic}
\newcommand{\Athens}{Department of Physics, University of Athens, 
Athens, 15771, Greece}
\newcommand{\BHU}{Department of Physics, Banaras 
Hindu University, Varanasi, 221 005, India}
\newcommand{\UCLA}{Physics and Astronomy Department, UCLA, Box 951547, Los 
Angeles, California 90095-1547, USA}
\newcommand{\Caltech}{California Institute of 
Technology, Pasadena, California 91125, USA}
\newcommand{\Cochin}{Department of Physics, Cochin University
of Science and Technology, Kochi 682 022, India}
\newcommand{\Charles}
{Charles University in Prague, Faculty of Mathematics and Physics,
 Institute of Particle and Nuclear Physics, Prague, Czech Republic}
\newcommand{\Cincinnati}{Department of Physics, University of Cincinnati, 
Cincinnati, Ohio 45221, USA}
\newcommand{\CTU}{Czech Technical University in Prague,
Brehova 7, 115 19 Prague 1, Czech Republic}
\newcommand{\Dallas}{Physics Department, University of Texas at Dallas,
800 W. Campbell Rd. Richardson, Texas 75083-0688, USA}
\newcommand{\Delhi}{Department of Physics \& Astrophysics, University of 
Delhi, Delhi 110007, India}
\newcommand{\JINR}{Joint Institute for Nuclear Research Joliot-Curie, 6 
Dubna, Moscow region 141980, Russia}
\newcommand{\FNAL}{Fermi National Accelerator Laboratory, Batavia, 
Illinois 60510, USA}
\newcommand{\UFG}{Instituto de F\'{i}sica, Universidade Federal de 
Goi\'{a}s, Goi\^{a}nia, GO, 74690-900, Brazil}
\newcommand{\Guwahati}{Department of Physics, IIT Guwahati, Guwahati, 781 
039, India}
\newcommand{\Harvard}{Department of Physics, Harvard University, 
Cambridge, Massachusetts 02138, USA}
\newcommand{\IHyderabad}{Department of Physics, IIT Hyderabad, Hyderabad, 
502 205, India}
\newcommand{\Hyderabad}{School of Physics, University of Hyderabad, 
Hyderabad, 500 046, India}
\newcommand{\Indiana}{Indiana University, Bloomington, Indiana 47405, 
USA}
\newcommand{\INR}{Inst. for Nuclear Research of Russian, Academy of 
Sciences 7a, 60th October Anniversary prospect, Moscow 117312, Russia}
\newcommand{\Iowa}{Department of Physics and Astronomy, Iowa State 
University, Ames, Iowa 50011, USA}
\newcommand{\Jammu}{Department of Physics and Electronics, University of 
Jammu, Jammu Tawi, 180 006, J\&K, India}
\newcommand{\Lebedev}{Nuclear Physics Department, Lebedev Physical 
Institute, Leninsky Prospect 53, 119991 Moscow, Russia}
\newcommand{\MSU}{Department of Physics \& Astronomy, Michigan State 
University, East Lansing, Michigan 48824, USA}
\newcommand{\Duluth}{Department of Physics \& Astronomy, 
University of Minnesota - Duluth, Duluth, Minnesota 55812, USA}
\newcommand{\Minnesota}{School of Physics and Astronomy, University of 
Minnesota - Twin Cities, Minneapolis, Minnesota 55455, USA}
\newcommand{\Oxford}{Subdepartment of Particle Physics, 
University of Oxford, Oxford OX1 3RH, United Kingdom}
\newcommand{\Panjab}{Department of Physics, Panjab University, 
Chandigarh, 106 014, India}
\newcommand{\RAL}{Rutherford Appleton Laboratory, Science and 
Technology Facilities Council, Didcot, OX11 0QX, United Kingdom}
\newcommand{\Carolina}{Department of Physics and Astronomy, University of 
South Carolina, Columbia, South Carolina 29208, USA}
\newcommand{\SDakota}{South Dakota School of Mines and Technology, Rapid 
City, South Dakota 57701, USA}
\newcommand{\SMU}{Department of Physics, Southern Methodist University, 
Dallas, Texas 75275, USA}
\newcommand{\Stanford}{Department of Physics, Stanford University, 
Stanford, California 94305, USA}
\newcommand{\Sussex}{Department of Physics and Astronomy, University of 
Sussex, Falmer, Brighton BN1 9QH, United Kingdom}
\newcommand{\Tennessee}{Department of Physics and Astronomy, 
University of Tennessee, 1408 Circle Drive, Knoxville, Tennessee 37996, USA}
\newcommand{\Texas}{Department of Physics, University of Texas at Austin, 
1 University Station C1600, Austin, Texas 78712, USA}
\newcommand{\Tufts}{Department of Physics and Astonomy, Tufts University, Medford, 
Massachusetts 02155, USA}
\newcommand{\Virginia}{Department of Physics, University of Virginia, 
Charlottesville, Virginia 22904, USA}
\newcommand{\WSU}{Physics Division, Wichita State Univ., 1845 
Fairmout St., Wichita, Kansas 67220, USA}
\newcommand{\WandM}{Department of Physics, College of William \& Mary, 
Williamsburg, Virginia 23187, USA}
\newcommand{\Winona}{Department of Physics, Winona State University, P.O. 
Box 5838, Winona, Minnesota 55987, USA}
\newcommand{\Crookston}{Math, Science and Technology Department, University 
of Minnesota -- Crookston, Crookston, Minnesota 56716, USA}
\newcommand{\deceased}{Deceased.}

\affiliation{\ANL}
\affiliation{\IOP}
\affiliation{\Athens}
\affiliation{\BHU}
\affiliation{\UCLA}
\affiliation{\Caltech}
\affiliation{\Charles}
\affiliation{\Cincinnati}
\affiliation{\Cochin}
%\affiliation{\CSU}
\affiliation{\CTU}
\affiliation{\Delhi}
\affiliation{\FNAL}
\affiliation{\UFG}
\affiliation{\Guwahati}
\affiliation{\Harvard}
\affiliation{\Hyderabad}
\affiliation{\IHyderabad}
\affiliation{\Indiana}
\affiliation{\INR}
\affiliation{\Iowa}
\affiliation{\Jammu}
\affiliation{\JINR}
\affiliation{\Lebedev}
\affiliation{\MSU}
\affiliation{\Crookston}
\affiliation{\Duluth}
\affiliation{\Minnesota}
\affiliation{\Oxford}
\affiliation{\Panjab}
\affiliation{\RAL}
\affiliation{\Carolina}
\affiliation{\SDakota}
\affiliation{\SMU}
\affiliation{\Stanford}
\affiliation{\Sussex}
\affiliation{\Tennessee}
\affiliation{\Texas}
\affiliation{\Dallas}
\affiliation{\Tufts}
\affiliation{\Virginia}
\affiliation{\WSU}
\affiliation{\WandM}
\affiliation{\Winona}

\author{P.~Adamson}
\affiliation{\FNAL}

%For first publication
\author{C.~Ader}
\affiliation{\FNAL}

%For first publication
\author{M.~Andrews}
\affiliation{\FNAL}

\author{N.~Anfimov}
\affiliation{\JINR}

\author{I.~Anghel}
\affiliation{\Iowa}
\affiliation{\ANL}

%For first publication
\author{K.~Arms}
\affiliation{\Minnesota}

\author{E.~Arrieta-Diaz}
\affiliation{\SMU}

\author{A.~Aurisano}
\affiliation{\Cincinnati}

%For first publication
\author{D.~Ayres}
\affiliation{\ANL}

\author{C.~Backhouse}
\affiliation{\Caltech}

% ...

\collaboration{The NOvA Collaboration}
\noaffiliation

\date{\today}
\preprint{FERMILAB-PUB-16-007-ND}

\begin{abstract}
This paper reports the first measurement using the \nova detectors of $\nu_\mu$ disappearance in a $\nu_\mu$ beam.  The analysis uses a 14 kton-equivalent exposure of \effpot from the Fermilab NuMI beam.  Assuming the normal neutrino mass hierarchy, we measure 
$\Delta m^{2}_{32}=(2.52\,^{+0.20}_{-0.18})\mathord{\times}10^{-3}$\,eV$^{2}$ and
$\sin^2\theta_{23}$ in the range 0.38--0.65, both at the 68\% confidence level, with two statistically-degenerate best fit points at $\sin^2\theta_{23}\,\mathord{=}\,0.43$ and 0.60.  Results for the inverted mass hierarchy are also presented.  
\end{abstract}

\maketitle

Neutrino oscillation is a powerful tool for probing fundamental neutrino properties~\cite{ref:SNO,ref:SuperKSolar,ref:kamland, ref:T2KCombined, ref:MinosCombined, ref:RENOLatest, ref:IceCube,ref:SuperKAtmos,ref:DoubleChooz, ref:DayaBay}. For the case of three-flavor mixing this process is governed by two independent mass-squared splittings, $\Delta m^{2}_{21}$ and $\Delta m^{2}_{32}$, and the unitary mixing matrix $U_\mathrm{PMNS}$~\cite{ref:PMNS}.  This matrix, which describes the linear combinations of neutrino mass eigenstates that constitute the neutrino flavor states, is parameterized by three angles $\theta_{13}$, $\theta_{23}$, and $\theta_{12}$,  and a CP-violating phase $\delta_{\rm CP}$.  $\theta_{23}$ has the largest measurement uncertainty of all mixing angles and is consistent with maximal mixing ($\theta_{23}\,\mathord{=}\,\pi/4$) within current experimental uncertainties~\cite{ref:DayaBay,ref:IceCube,ref:T2KCombined,ref:MinosCombined,ref:SuperKAtmos}.
%\cite{ref:IceCube,ref:SuperKAtmos,ref:T2KCombined, ref:MinosCombined}.
The observation of $\nu_\mu$ disappearance, as reported here, is used to measure $\Delta m^{2}_{32}$ and $\sin^2\theta_{23}$.  Precise knowledge of $\theta_{23}$ is an input into future $\nu_e$ and $\overline{\nu}_e$ appearance measurements that may determine whether $\nu_1$ or $\nu_3$ is the lightest mass eigenstate (normal or inverted mass hierarchy, respectively), whether 
$\theta_{23}\,\mathord{>}\,\pi/4$ or $\theta_{23}\,\mathord{<}\,\pi/4$, and whether neutrinos violate CP symmetry.  This paper reports the first measurement by the \nova experiment of $\sin^2\theta_{23}$ and $\Delta m^{2}_{32}$ via $\nu_\mu$ disappearance.% in a $\nu_\mu$ beam.

%All of these measurements require precise knowledge of $\theta_{23}$.  %Furthermore, a determination that $\theta_{23}$ mixing is not maximal could perhaps be a first step towards a measurement of the sign of $\theta_{23}-\pi/4$, and of the study of broken symmetries of the PMNS matrix. 
%The determination of the sign of $\Delta m^{2}_{32}$, for example, will answer whether $\nu_1$ or $\nu_3$ is the lightest neutrino mass eigenstate.  A non-zero value of $\sin\delta_{\rm CP}$ indicates that neutrinos violate CP symmetry, which could have resulted in baryogenesis in the early universe and thus explain the matter-antimatter asymmetry.  The resolution of both these issues will most likely be based on f

Neutrinos produced in the NuMI beamline at Fermilab~\cite{ref:NuMI} are observed in the \nova Near Detector (ND) on the Fermilab site and the \nova Far Detector (FD) 810\,km from the NuMI target along the Ash River Trail, MN~\cite{ref:nova}. The 14-kton FD is positioned on the surface, 14.6\,mrad off-axis from the NuMI beam. The 290-ton ND, 100\,m underground and 1\,km from the NuMI target, is also positioned off-axis to allow a measurement of an unoscillated neutrino energy spectrum that closely matches the unoscillated spectrum at the FD. The kinematics of two-body $\pi \rightarrow \mu + \nu_\mu$ decay in the NuMI decay pipe results in a neutrino energy spectrum in the off-axis detectors that peaks close to 2\,GeV, near the first maximum of the \numu disappearance probability at the FD.  The neutrino energy spectrum has a FWHM of approximately 1\,GeV.

The \nova detectors are functionally identical, segmented, tracking calorimeters.  The detectors are designed to provide sufficient sampling of hadronic and electromagnetic showers to allow efficient separation of the \nue and \numu~charged current (CC) interaction signals from the neutral current (NC) interaction backgrounds.  The basic unit of the \nova detector is a long liquid-scintillator-filled cell with highly reflective white polyvinyl chloride (PVC) walls and cross sectional size 3.9\,cm by 6.6\,cm.  The liquid scintillator comprises 62\% of the fiducial mass of each detector, and a minimum ionizing particle deposits approximately 1.8\,MeV of energy for each centimeter traveled in the scintillator of each cell.  The PVC cells have a length of 15.5\,m in the FD, and 3.9\,m in the ND.  Each cell contains a Kuraray Y11 wavelength-shifting 0.7\,mm diameter fiber~\cite{ref:Kuraray} that runs the length of a cell, loops, and returns to the readout end where both ends of the fiber terminate on a single pixel of a Hamamatsu avalanche photodiode (APD)~\cite{ref:Hamamatsu} operated in proportional gain mode. 

Planes of PVC cells with their long-axes alternating between horizontal and vertical orientations allow three-dimensional reconstruction of tracks and showers.  The FD consists of 896\,planes of 384\,cells each and is 59.8\,m in length.  The ND is 15.3\,m in length and consists of 192 contiguous upstream PVC planes with 96\,cells each.  At the downstream end of the ND a muon range stack is formed of 11 pairs of active vertical and horizontal PVC planes, with a 10~cm thick steel plane between each pair.  The muon range stack is two-thirds the height of the bulk ND and thus the active horizontal planes have 64 cells rather than 96.  The muon range stack is used to improve the containment of muons produced in the upstream active volume of the detector.  
%The longitudinal axis of each detector is aligned to the average beam direction at their respective locations.  

The digitization and processing of APD signals is continuous and deadtime-free.  The signals produce pulse-height and timing information for any signal above a  pulse-height threshold corresponding to approximately 75\% of that expected for the passage of a minimum ionizing particle through a detector cell at the end furthest from the APD in the FD.   Data are recorded in 550\,\mus-long trigger windows roughly centered on the 
10\,\mus-long NuMI spills.  Additional trigger windows are taken out of time with the beam spill to collect cosmic ray events for calibration and background studies. 
 
%The timing resolution of the recorded data is approximately 140\,ns.

% by a front-end board (FEB) containing all of the electronics needed to shape and analyze %the signal.  Each APD package is operated at a gain of 100 and cooled to a temperature of -15$^\circ$ C %to achieve a quantum efficiency of approximately 80\% across the entire spectrum of light transmitted by %the fiber.  %As indicated in the photo on the left of Fig.~\ref{fig:FEBAPD}, t
%The FEB contains 32 amplifiers with shaping circuitry in an application-specific integrated circuit %(ASIC), four-fold analog-to-digital converters (ADCs), a field-programmable gate array (FPGA) and a %thermo-electric cooler controller (TECC).  The TECC is used to maintain the operating temperature of the %APD.  The ASIC and ADCs shape and convert the analog signals from the APD to digital signals that are %time-stamped and formatted by the FPGA which performs a pixel-by-pixel comparison to programmable %thresholds.  The thresholds are determined from dedicated measurements of the baseline response to noise %in each pixel.  The FPGA sends its zero-suppressed data to a Data Concentrator Module (DCM).  A DCM %collects data from up to 64 FEBs.  Data collected in each DCM are time-sorted and sent to a computing %farm where several minutes of data from all DCMs are buffered.  %An external trigger or a local data-%driven trigger result in the permanent recording of the buffered data.  

The neutrino beam used in this study is generated by colliding 120\,GeV protons from the Fermilab Main Injector onto a 1.2\,m graphite target.  Two magnetic horns located downstream of the target %are pulsed with a 200\,kA current in time with the Main Injector beam spill, yielding a maximum 30\,T toroidal field which 
focus charged particles of one sign along the beam direction and defocus charged particles of the opposite sign.  %The magnetic horns are followed by a 675~m long, 2~m diameter pipe filled with He at 1~atm in which the neutrino beam is produced via meson and muon decay.  
%A hadron monitor and absorber is located at the end of the decay pipe, followed by 240 m of rock to range out muons, resulting in an essentially pure neutrino beam.  
With the horns focusing positive mesons, simulations predict that the \nova off-axis detectors are exposed to a neutrino beam composed of 97.6\% $\nu_\mu$, 1.7\% $\overline{\nu}_\mu$ and 0.7 $\nu_e + \overline{\nu}_e$ for neutrino energies between 1 and 3\,GeV.  

In both detectors we measure the energy spectrum of muon-neutrino CC interactions, primarily on carbon nuclei.  The flavor and energy of the incident neutrino is determined by identifying the lepton flavor in the final state and assigning all the energy deposited by the final state particles to the neutrino.  The measured FD neutrino spectrum is fit to a predicted spectrum based on measurements of the unoscillated spectrum in the ND and the effect of neutrino oscillations.  Monte Carlo simulations are used to correct for beam flux and acceptance differences between the two detectors.

%Overview of the MC

%The simulated data used in this analysis are generated and reconstructed with detector configurations that match the real data configuration.  
The simulation of the neutrino flux produced by the NuMI beamline is based on {\small FLUGG}~\cite{ref:FLUGG} which uses the 
{\small FLUKA}~\cite{ref:FLUKA} and {\small GEANT4}~\cite{ref:GEANT4} simulations. 
%FLUGG 2009.4~\cite{FLUGG} event generator
It includes a full simulation of the production of hadrons by the 120\,GeV primary proton beam interacting with the NuMI target and the propagation of those hadrons through the target, magnetic horns, and along the decay pipe.  
%Already have cite for flugg so you don't need to go into more detail
%FLUGG is a combination of {\small FLUKA}~\cite{ref:FLUKA} and {\small GEANT4}~\cite{ref:GEANT4}. {\small FLUKA} simulates the scattering of protons on the graphite target and interaction of secondary particles within the target and stores the kinematic and particle type information of produced particles as they exit the surface of the NuMI target.  The output of the {\small FLUKA} simulation is the input to a {\small GEANT4} simulation which propagates particles through the magnetic horns and NuMI decay pipe. 
The generation of neutrino interactions in the \nova detector and surrounding rock is performed using the {\small GENIE} simulation~\cite{ref:GENIE}.  
%GENIE uses the output flux from FLUGG simulation and predicted neutrino cross sections to determine whether or not individual neutrinos interact in the detector material.  
{\small GENIE} simulates the primary interaction inside the nucleus, the production of all final-state particles in the nucleus (hadronization), and the transport and rescattering of the final-state particles through the nucleus (intranuclear transport).  For this analysis, three charged-current neutrino interaction types categorized by {\small GENIE} dominate the signal: quasielastic (QE), baryon resonance production (RES) and deep-inelastic scattering (DIS).  %In QE interactions the neutrino scatters on a bound neutron and ejects a proton (and possibly other nucleons as well) from the target. This scattering is simulated using the Llewellyn Smith formulation of weak interaction V-A phenomenology~\cite{Smith}. The RES events are baryon resonance production events whose decay yields a single pion and a nucleon in the final state, and are simulated using the Rein-Sehgal model~\cite{ReinSehgal}.  In DIS interactions, the neutrino scatters off a quark and produces a lepton and a hadronic system in the final state.  DIS events are simulated using an effective leading-order model by Bodek and Yang~\cite{BodekYang}.  
%NC interactions have no charged lepton in the final state.  
The transport, energy loss, interactions and decays of final state particles within the detector volume are simulated by {\small GEANT4}.  The {\small GEANT4} simulation uses a description of the geometry and material content of the detectors.  The simulated energy deposition in the liquid scintillator is converted to a corresponding number of photoelectrons observed in the APD using a model of light production, capture and propagation in the fiber that is based on test-stand measurements.  This photoelectron signal is then converted to digitized quantities in the same format as data collected from the detectors using a model of the readout electronics response, also based on test-stand measurements.

The FD data used in this analysis come from an exposure of \realpot~(POT). This includes periods during FD construction when a fraction of the detector was live. On average 79.4\% of the detector was live over the data set, which corresponds to a full FD 14 kton-equivalent exposure of \effpotPOT.  The varying size of the FD is accounted for in the simulation.  In addition, data collection in the ND began later than in the FD.  The resulting ND data sample, which was recorded with a fully instrumented detector, corresponds to \ndpot~POT. 

%A blind analysis is conducted such that the FD neutrino spectrum is only determined after all FD data selection criteria are established and systematic uncertainties assessed.  Once the FD neutrino spectrum is unblinded, the data are fit to a predicted spectrum based on measurements of the unoscillated spectrum in the ND and Monte Carlo (MC) simulations to correct for flux and acceptance differences between the two detectors.    
%Leave o ut this information since plots can show the width %The kinematics of the off-axis location of the detectors results in a 800 MeV full-width half-maximum narrow-band beam with the neutrino energy spectrum peaked near 2~GeV.  %The peak position corresponds to the first oscillation maximum for $\nu_\mu \rightarrow \nu_e$ oscillations at a baseline of 810 km.  
%The data used in this analysis were collected from an exposure of \realpot while the detectors were under construction.  After the commissioning of sections of far detector (FD) were completed, the detector sections were included in the analysis readout.  

The energy response of each channel in the detector is individually calibrated using cosmic-ray muons.  The observed signals for muon energy depositions at different distances along the length of each cell are used to characterize the signal attenuation in the fiber of that cell.   %The conversion of the attenuation-corrected channel response to cell energy deposition is accomplished using the observed response to hits located 100-200 cm from the end of the range of muons stopping in the detector, the range in which the cosmic ray muons are minimum ionizing.

\begin{figure}[!htb]
\begin{center}
\includegraphics[width = 0.49\textwidth]{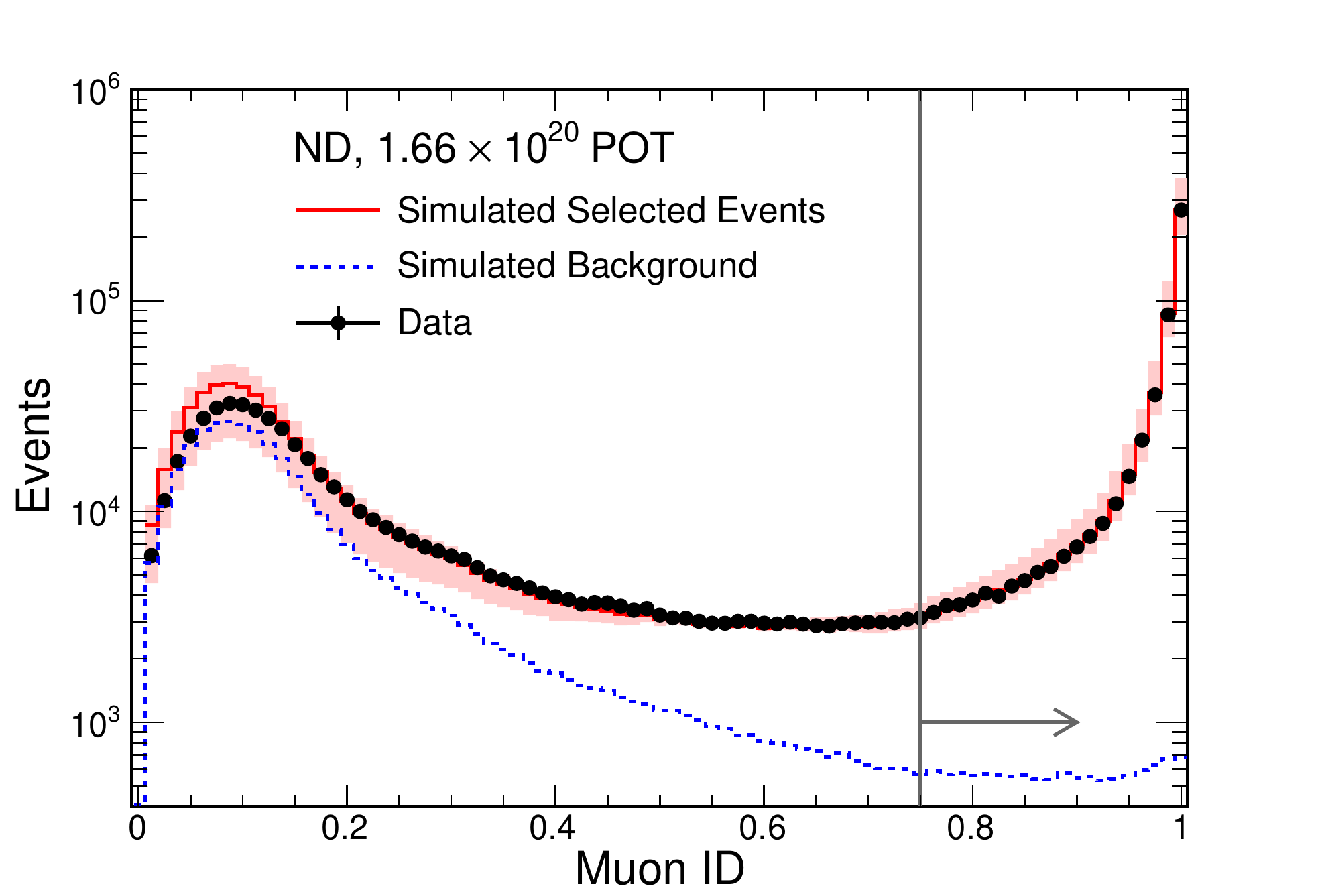}
\end{center}
\caption{The muon identification variable in the ND for contained neutrino events. For each event only the largest muon ID of all reconstructed tracks is shown. Events with muon ID greater than 0.75 are selected for analysis.  The simulated distribution (solid red) and its background component (dashed blue) are also shown.  The simulated distribution is normalized up for display purposes to remove a 7.2\% offset after selection criteria are applied.  The shaded band represents the bin-to-bin uncertainties only, suppressing the 20-30\% normalization uncertainties due primarily to neutrino flux and cross sections.}
\label{fig:ReMId}
\end{figure}

Event reconstruction and characterization starts from calibrated cell data that are grouped into collections based on their proximity in both space and time~\cite{ref:DBSCAN,ref:Baird}.  The cell data in each collection is assumed to arise from the same primary neutrino or cosmic-ray event.  The 
cell energy depositions in these events are then used to reconstruct charged particle trajectories.  In this analysis the reconstruction of muon tracks produced in \numu~CC interactions is performed using an algorithm based on the Kalman filter technique~\cite{nickdpf,Kalman}.

%A series of containment selection criteria serves to ensure that all of the energy from the neutrino interaction is deposited within the detector, to reject muons produced by neutrino interactions in the rock surrounding the detectors, and to reduce the cosmic ray background in the FD.  The selection criteria require no hits in the outermost two cells and planes.  Reconstructed primary tracks that would require fewer than 4 cells to reach the edge of the detector along the forward projection of the track from the end of the track, and fewer than 8 cells along the backward projection from the beginning of the track, are also rejected in the ND.  A more stringent projection containment of 10 cells in both the forward and backward directions is applied in the FD.  In the ND, both the interaction vertex and all energy deposited in hits not associated with the selected muon are required to be upstream of the muon range stack.  %These requirements optimize the simulation-based reconstructed energy resolution of \numu CC interactions in each detector.

\begin{figure*}[!tbp]
   \centering
   \includegraphics[width=0.49\textwidth]{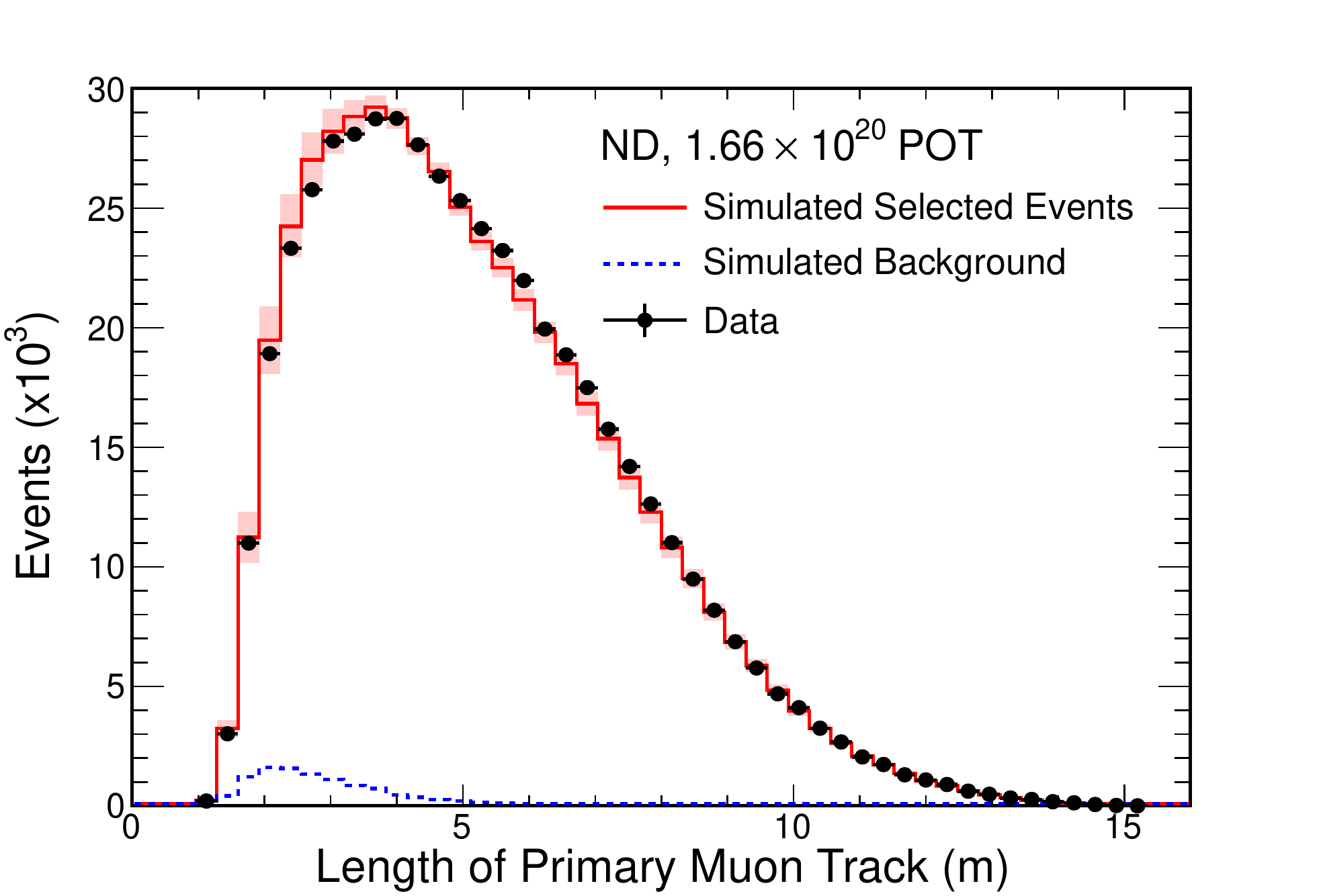} 
   \includegraphics[width=0.49\textwidth]{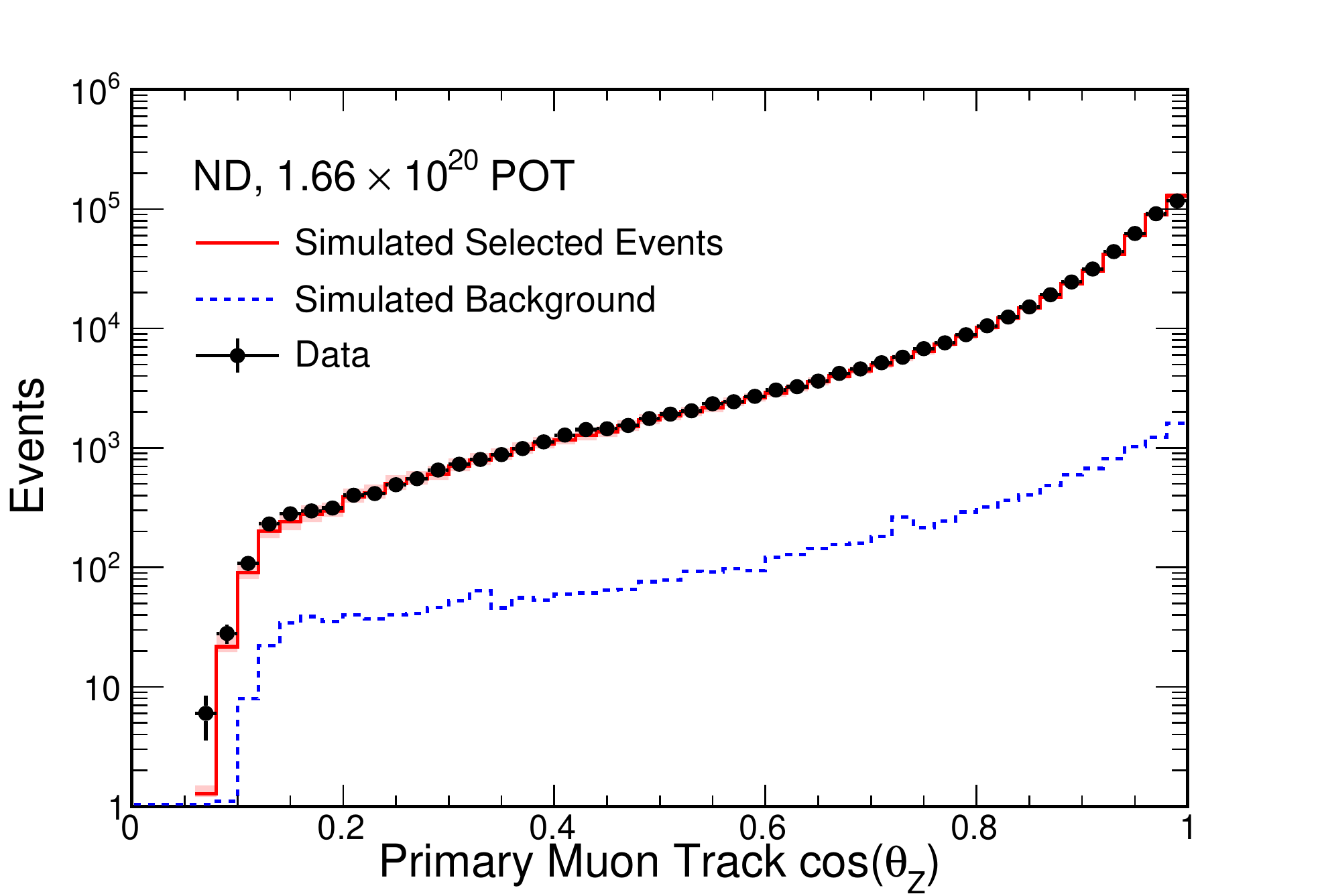}
   \caption{Reconstructed track length (left) and track angle $\theta_Z$ relative to the detector longitudinal axis, along the beam direction (right) for the primary muons in selected \numu~CC interactions in the ND.  The simulated distributions follow the conventions of Fig.~\ref{fig:ReMId}.}
   \label{fig:NDDists}
\end{figure*}

%!!!!!!!!!!!!!!!!!
A multivariate analysis implementing a $k$-Nearest Neighbor algorithm~\cite{nickdpf,ref:kNN,ref:Rustem} is used to identify a muon track in the reconstructed event.  The resulting muon identification (ID) is based on the measured $dE/dx$, amount of multiple scattering along the track, total track length, and the fraction of track planes that have overlapping hadronic activity.  The muon identifier was trained separately for each detector using simulated \numu interactions. For events with multiple tracks, the primary muon candidate is the track with the highest muon ID. Events are selected as \numu~CC if the primary track has a muon ID score greater than 0.75. The distribution of the muon identification variable for the primary tracks of contained ND data and simulated neutrino events is shown in Fig.~\ref{fig:ReMId}.%(the simulated distribution in the FD has a similar shape and is not shown). 

The reconstructed neutrino events are required to be fully contained in the detectors to ensure an accurate measurement of the neutrino energy, to reject muons produced by neutrino interactions in the rock surrounding the detectors, and to reduce the cosmic-ray background in the FD.  In order to contain hadronic activity, the selection criteria require that the event has no energy depositions in the two cells and planes that are nearest to the detector edge.  To ensure that the muon is contained, requirements are placed on the start and end positions of the primary track.  In the ND, the forward projection of the track must be 4 or more cell-widths away from the edge of the detector and the backward projection of the track must be 8 or more cell-widths away from the detector edge.  A more stringent projection requirement of 10~cells in both the forward and backward directions is applied in the FD, due to the larger cosmic-ray rate for the surface detector.  To ensure that the energy resolution in the ND is comparable to the FD, an additional containment requirement is applied in which both the interaction vertex and all but 30\,MeV of energy deposited in cells not associated with the selected muon must be upstream of the muon range stack.  %These requirements optimize the simulation-based reconstructed energy resolution of \numu CC interactions in each detector.

%During the collection of the data presented here, the NuMI beam line operated at a typical intensity of $2.5 \times 10^{13}$ POT per spill, with a spill duration of 10 \mus delivered every 1.3~s.  The FD is located on the surface, with an overburden of 1.3~m of concrete and 15~cm of barite (total of 3 meters-water-equivalent), and records a background rate of 148~kHz of reconstructed cosmic ray-induced events.  These should be moved to earlier sections
The rate of reconstructed cosmic ray-induced events in the FD is 148\,kHz.  The corresponding background within the 10\,\mus beam window, mostly muons, is reduced using criteria determined from the high-statistics out-of-spill-time data sample and from simulated neutrino interactions.  %The selection of reconstructed events within the known timing of the NuMI beam spill window reduces the cosmic ray background by a factor of $10^5$.  
The event containment and muon identification criteria described above reduce this background rate in the FD by a factor of approximately 200.  Additional selection criteria based on the primary track angle, which is generally beam-directed for neutrino-produced muons and downwards-directed for cosmic rays, as well as the number of energy deposits in cells in the event, further reduce the background by two orders-of-magnitude.  A final three orders-of-magnitude in background rejection, removing the most signal-like cosmic rays, is achieved with a boosted decision tree.  This multivariate algorithm utilizes eleven variables, based on the reconstructed tracks (direction, multiple scattering, length, number of tracks, and fraction of cells with energy deposits associated to the muon track), event calorimetry, and general event topology (proximity to detector top and edges).

Approximately 57\% of simulated contained \numu~CC events with less than 5.0\,GeV of visible energy pass all of the FD selection criteria, whereas the cosmic background with visible energy below 5.0\,GeV is reduced by a factor of $1.2\times 10^{7}$. With this level of rejection the cosmic background contributes 4.1\% of selected FD \numu~CC events. The uncertainty on the cosmic background was determined using the out-of-spill data and is negligible for this analysis. The background from contained NC events within the same visible energy range is estimated using simulation to contribute 6\% of selected FD \numu~CC events, which is a 99\% reduction. The \nue and \nutau~CC interactions are negligible backgrounds in both detectors. %can't put backgrounds in terms of % for FD since total # of FD events depends on osc pars --- except we do have a fixed number of events in FD data... so use that.

%Approximately 57\% of simulated contained \numu~CC events with less than 5.0~GeV of visible energy pass all of the FD selection criteria, whereas the cosmic background with visible energy below 5.0~GeV is reduced by a factor of $1.2\times 10^{7}$, resulting in an expected cosmic background of 1.4$\pm$0.2 events in the final data sample.  The background from contained NC events within the same visible energy range is reduced by 98.7\% to 2.0 events (with negligible statistical uncertainty).  The \nue and \nutau~CC interactions are negligible backgrounds in both detectors. %can't put backgrounds in terms of % for FD since total # of FD events depends on osc pars

\begin{figure}[!htbp]
   \centering
   \includegraphics[width=0.49\textwidth]{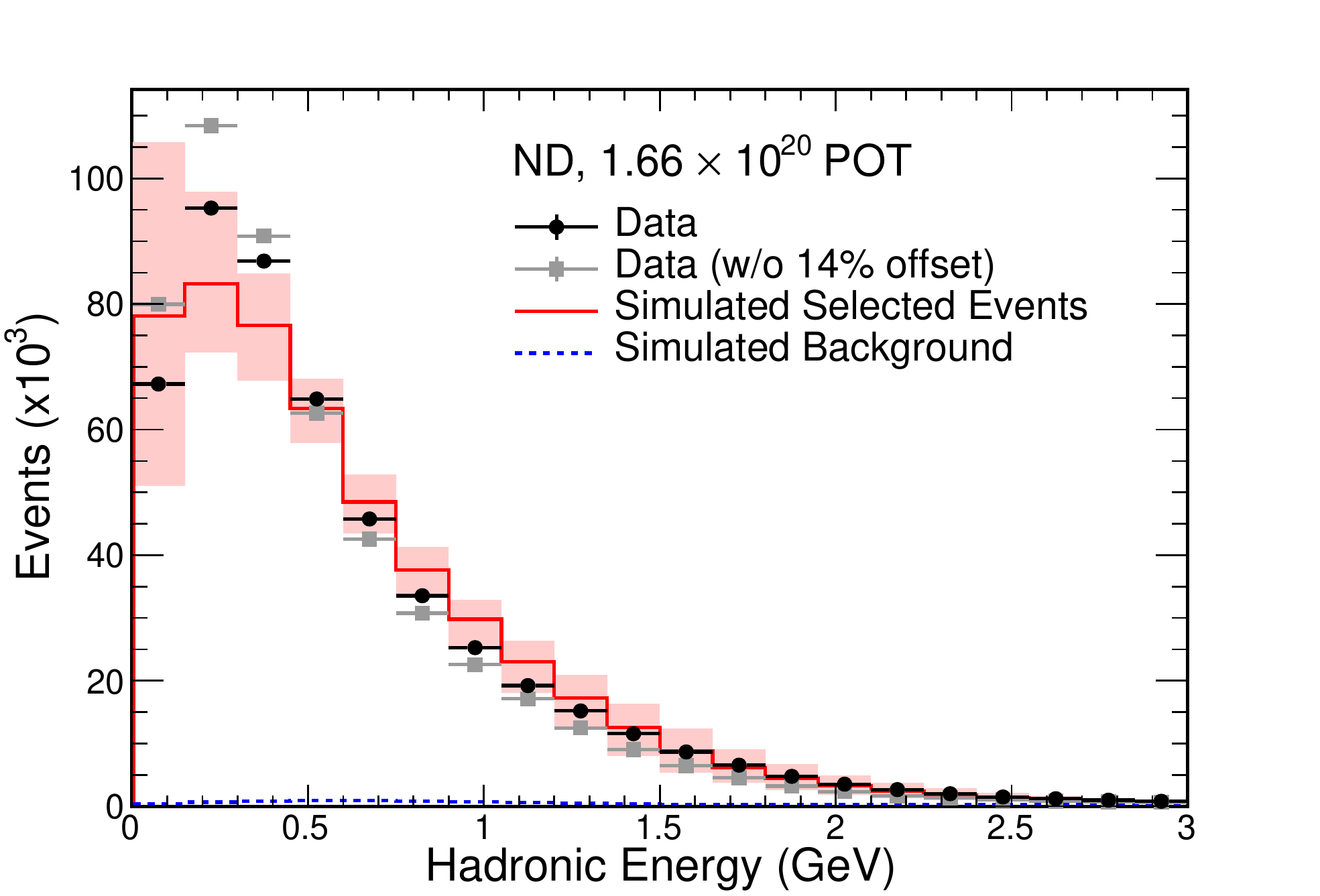} 
   \caption{Reconstructed hadronic energy \Ehad for selected \numu~CC interactions in the ND, both with (black circles) and without (gray squares) the 14\% offset described in the text.  The simulated distributions follow the conventions of Fig.~\ref{fig:ReMId}.}
   \label{fig:NDEHad}
\end{figure}

In the ND, the selected sample is estimated by simulation to be 98\% pure, with 2\% NC contamination.  Since the ND is underground, the cosmic-ray background is negligible. Backgrounds from muons produced by neutrino interactions in the surrounding rock are also negligible after containment requirements are applied. %Only 4\% of all \numu~CC interactions with vertices in the detector are selected, due to the low geometric acceptance of the ND. 
After all selection criteria, approximately 500,000 events remain in the ND data sample.  

%The efficiency of reconstructing and identifying muon tracks in the \nova detectors is extremely high, %, whereas the 
%current event reconstruction has a low efficiency for reconstructing low energy hadrons.  
The reconstructed neutrino energy $E_{\nu}$ of a contained \numu~CC event is given by 
\begin{equation}
E_{\nu} = E_\mu + E_\mathrm{had},
\end{equation}
where $E_\mu$ is the estimated energy of the primary muon track based on its reconstructed path length through the detector and \Ehad is the estimated energy of the hadronic shower based on the sum of all calibrated energy deposition in the event not attributed to the muon~\cite{ref:susanThesis}.  To achieve better $E_{\nu}$ agreement between data and simulation in the ND, the \Ehad calibration scale in data is set 14\% higher than that for simulation.

\begin{figure}[!htbp]
   \centering
   \includegraphics[width=0.49\textwidth]{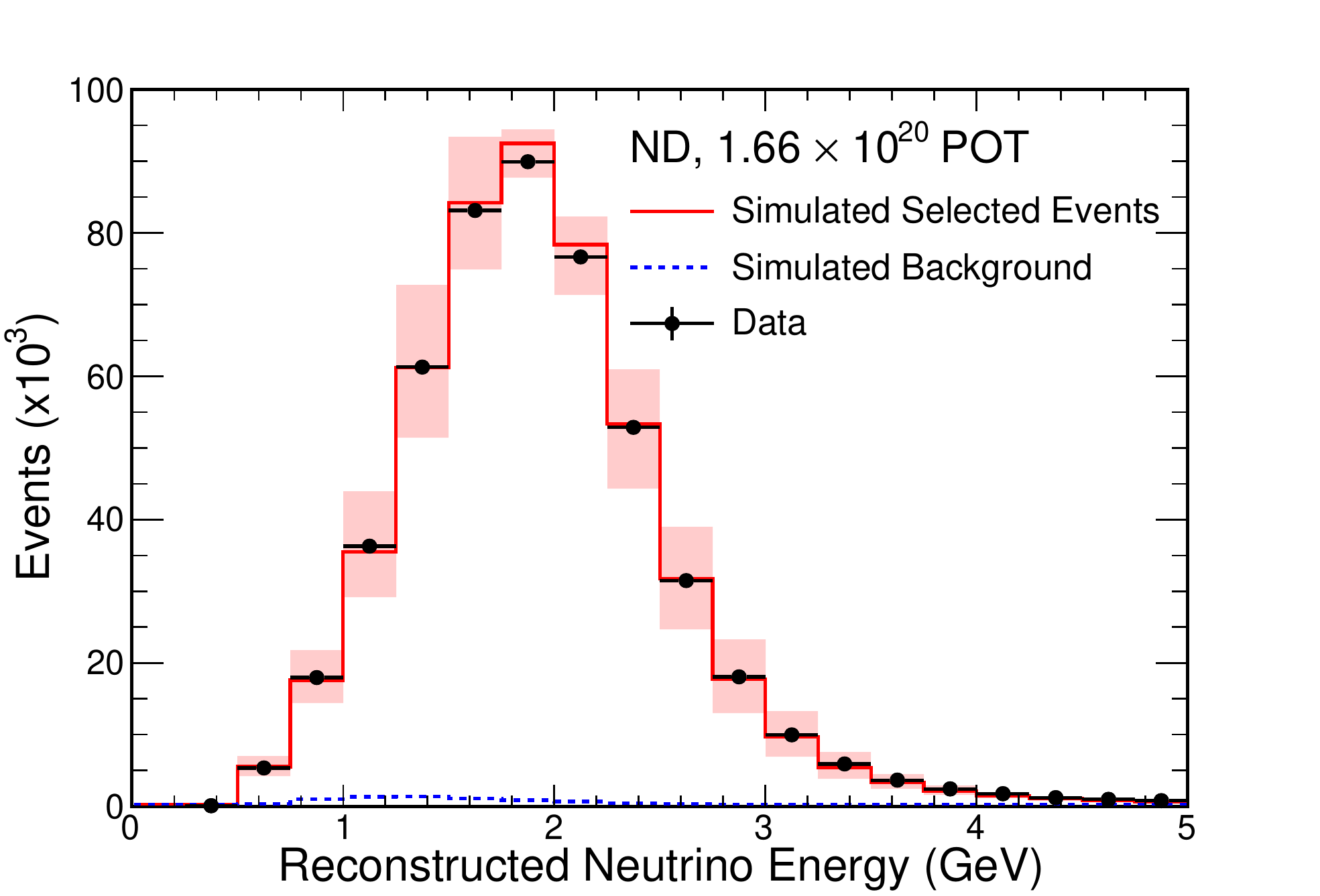}
   \caption{Reconstructed neutrino energy $E_{\nu}$ for selected \numu~CC interactions in the ND.  The simulated distributions follow the conventions of Fig.~\ref{fig:ReMId}.}
   \label{fig:NDSpectrum}
\end{figure}

Figure~\ref{fig:NDDists} shows the reconstructed muon track parameters for \numu{} CC events in the ND.  Figure~\ref{fig:NDEHad} shows the \Ehad{} distribution both with and without the 14\% difference in \Ehad calibration scale between data and simulation.  A corresponding $\pm14\%$ uncertainty is assessed on the hadronic energy scale, and is included in all of the uncertainty bands shown.  Figure ~\ref{fig:NDSpectrum} shows the final $E_{\nu}$ distribution.  The energy resolution for reconstructed \numu~CC events is estimated from simulation to be 7\%.

The prediction for the FD neutrino energy spectrum is based on the observed ND neutrino energy spectrum, with corrections for acceptance and flux differences derived from simulation.  First, the small NC background, estimated from simulation, is subtracted from the ND data spectrum.  The resulting background-subtracted spectrum is then converted into a true neutrino energy spectrum via a mapping derived from simulation. This true neutrino energy spectrum is then used to construct a spectrum in the FD by multiplying it by the energy-dependent ratio of FD-to-ND selected events from simulation.  Oscillation probabilities for a given set of oscillation parameters are then applied, by energy bin, to the predicted true FD energy spectrum, which  is  then mapped to  a reconstructed neutrino energy spectrum using FD simulation.  The extrapolated \numu~CC energy spectrum is then combined with beam-induced backgrounds (NC, \nue~CC, and \nutau~CC) predicted from simulation, and the background spectrum measured using events selected from outside of the beam spill window.

%Systematics:
Systematic uncertainties in the calibration, flux estimate, cross sections, hadronization modeling, particle-transport modeling and exposure differences between the two detectors are assessed by varying these aspects of the simulation.  Because the detectors are functionally identical, many systematic uncertainties largely cancel in the measurements of $\sin^2\theta_{23}$ and $\Delta m^2_{32}$.  The uncertainties assessed and their impact are summarized in Table~\ref{tab:errors}.

For the beam-induced backgrounds, which are small, a normalization uncertainty of 100\% is assigned.  The measured background outside the beam spill window has negligible uncertainty.  The neutrino interaction cross section and hadronization uncertainties are determined by altering each cross section and hadronization parameter by its predetermined uncertainties in the {\small GENIE} simulation, which vary in size from 15\% to 25\%, as specified in Ref.~\cite{ref:GENIEManual}.  Uncertainties in particle-transport modeling are assessed by comparing alternative hadronic models in the GEANT4 simulation.  The beam flux normalization uncertainty in each detector is dominated by beamline hadron production uncertainties.  This uncertainty is approximately 20\% near the peak of the spectrum, estimated by comparing simulated pion and kaon yields in the NuMI target to measured yields for interactions of 158\,GeV protons on a thin carbon target in the NA49 experiment~\cite{ref:NA49, ref:NA49K}.  The detector exposure uncertainty, which accounts for uncertainties in detector mass and periods of data collection when only one detector was operational, is 1\%. 

The uncertainty in muon energy scale is 2\%, driven by detector mass and muon energy-loss modeling.  The uncertainty in calorimetric (hadronic) energy scale is 14.9\%, the quadrature sum of the 14\% uncertainty assigned to reflect the difference in \Ehad{} scales used in data and simulation, and 5\% derived from comparisons of muon and Michel electron data and simulation.  An additional relative 5.2\% calorimetric energy uncertainty is taken uncorrelated between the two detectors.  The main component of this is a 5\% uncertainty derived from muon and Michel electron studies.  An additional 1.4\% comes from potential differences in \Ehad{} scale between the ND and FD due to their differing neutrino spectra (primarily due to oscillations).  To estimate this uncertainty, the simulated ND kinematic distributions were fit to data by adjusting some or all of the normalizations, hadronic energy scales, and muon energy scales of QE, RES, and DIS events separately in the simulation.  The fit results were then applied to FD simulation, and the largest relative energy offset seen between detectors across the ensemble of fits was 1.4\%.  The largest normalization offset seen was 1\%, which is also taken as an uncertainty.

\begin{table*}[t]
\begin{tabular}{c c c}
\hline 
\hline 
Source of Uncertainty & Fractional Uncertainty & Fractional Uncertainty \\
 & $\sin^2\theta_{23}$ $(\pm \%)$ & $\Delta m^2_{32}$ $(\pm \%)$ \\
\hline 
Absolute Calorimetric Energy Calibration ($14.9\%$) & 4.1 & 2.6 \\
Relative Calorimetric Energy Calibration ($5.2\%$) &  3.4 &  0.6 \\%0.92 \\
Muon Energy Scale ($2\%$) & 2.2 & 0.8\\
Cross Sections and Final State Interactions ($15-25\%$) & 0.8 & 0.6 \\ %0.55 & 0.83 \\
NC and \nutau~CC Backgrounds ($100\%$) & 3.0 & 0.6 \\ %0.87 \\
Particle-Transport Modeling & 1.5 & 0.6 \\ % 0.89 \\
Beam Flux ($21\%$) & 1.3 & 0.3 \\ % 0.63 \\
Normalization ($1.4\%$)& 0.4 & 0.2 \\ % 0.44 & 0.58 \\
Other Oscillation Parameters & 1.8 & 2.2 \\
\hline 
Total Systematic Uncertainty & 6.8 & 3.7 \\
\hline
Statistical	Uncertainty & 17.0 & 4.5 \\
\hline
\hline
\end{tabular}
\caption{Impact of the sources of uncertainty on the expected sensitivity of the measured values for $\sin^2\theta_{23}$ and $\Delta m^2_{32}$ evaluated at the test point of $\sin^2\theta_{23} = 0.5$ and $\Delta m^2_{32}=2.5 \times 10^{-3}$\,eV$^2$.}%The uncertainty on the absolute hadronic energy correction, which is an uncertainty in the energy calibration scale, dominates the systematic uncertainty on the measurement of $\sin^2\theta_{23}$, whereas the uncertainty on all other oscillation parameters dominates the systematic uncertainty on the measurement of $\Delta m^2_{32}$.}
\label{tab:errors}
\end{table*}

\begin{figure}[tb]
\centering
%\hfill 
\includegraphics[width=0.49\textwidth]{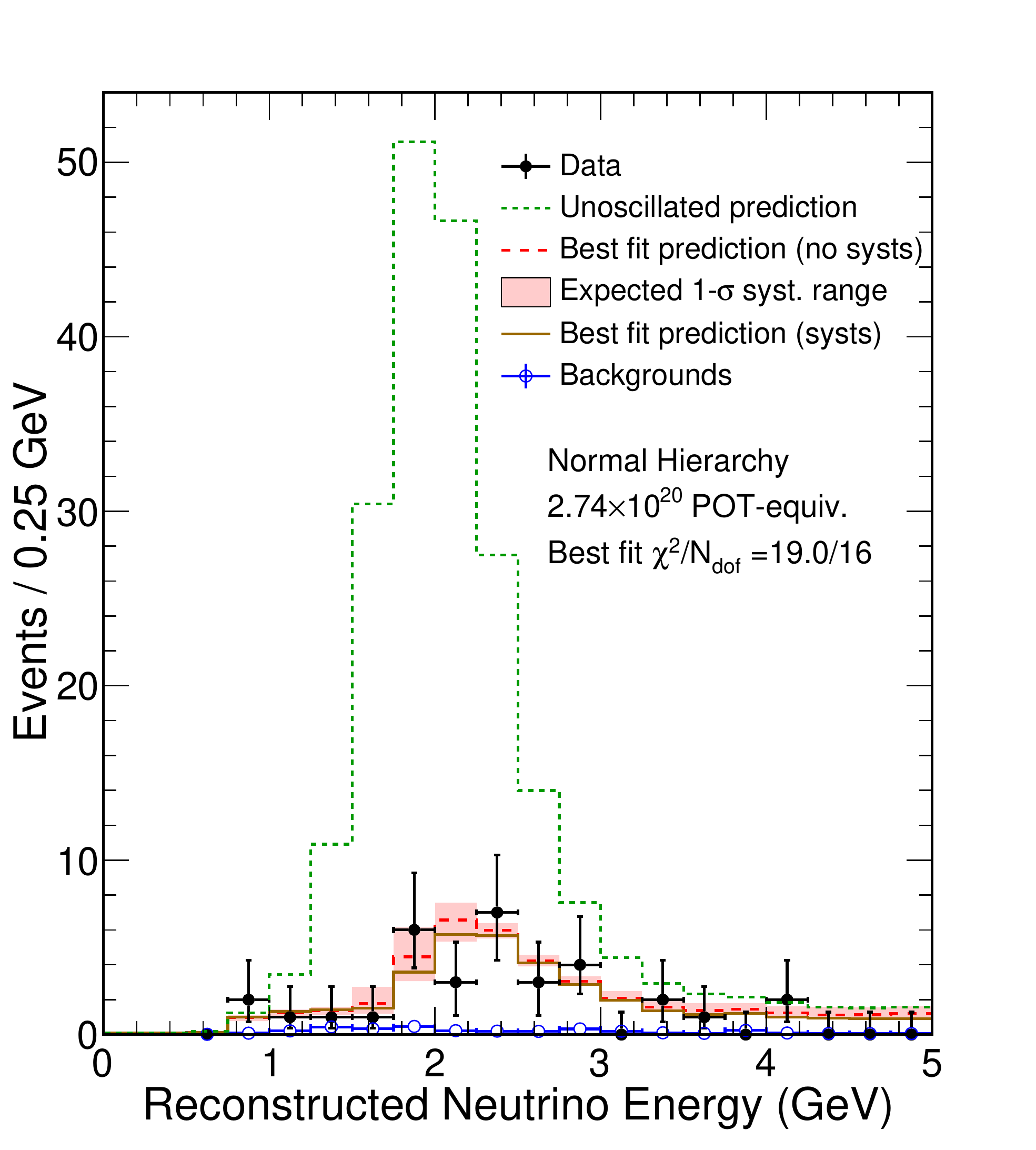}
\caption{The reconstructed energy for FD selected events. The black data points show the statistical uncertainties.  The short-dashed green histogram corresponds to the predicted spectrum in the absence of oscillations.  The solid brown histogram corresponds to the best fit prediction with systematic effects included.  The long-dashed red histogram corresponds to the best-fit prediction when the effects from the systematic shifts in the fit are removed.  The light-red band represents the systematic uncertainty on the no-systematics (red) prediction.  The blue, open-circled points represent the background, mostly NC and cosmic-ray muons.}
\label{fig:fdE}
\end{figure}

%FD Events:
Upon applying the FD event selection criteria to the full data set reported here, a total of 33 \numu~CC candidate events are observed for reconstructed neutrino energies below 5\,GeV. The total expected background is 3.4 events, which includes $2.0\,\mathord{\pm}\, 2.0$ NC events and $1.4\,\mathord{\pm}\,0.2$ cosmic-ray events. In the absence of neutrino oscillations $211.8\,\mathord{\pm}\,12.5$ (syst.)\ candidate events are predicted.  The energy spectrum for the sample is shown in Fig.~\ref{fig:fdE}.

%Approximately 57\% of simulated contained \numu~CC events with less than 5.0~GeV of visible energy pass all of the FD selection criteria, whereas the cosmic background with visible energy below 5.0~GeV is reduced by a factor of $1.2\times 10^{7}$, resulting in an expected cosmic background of 1.4$\pm$0.2 events in the final data sample.  The background from contained NC events within the same visible energy range is reduced by 98.7\% to 2.0 events (with negligible statistical uncertainty).  The \nue and \nutau~CC interactions are negligible backgrounds in both detectors. %can't put backgrounds in terms of % for FD since total # of FD events depends on osc pars

\begin{figure}[!htb]
\centering
%\hfill 
\includegraphics[width=0.49\textwidth]{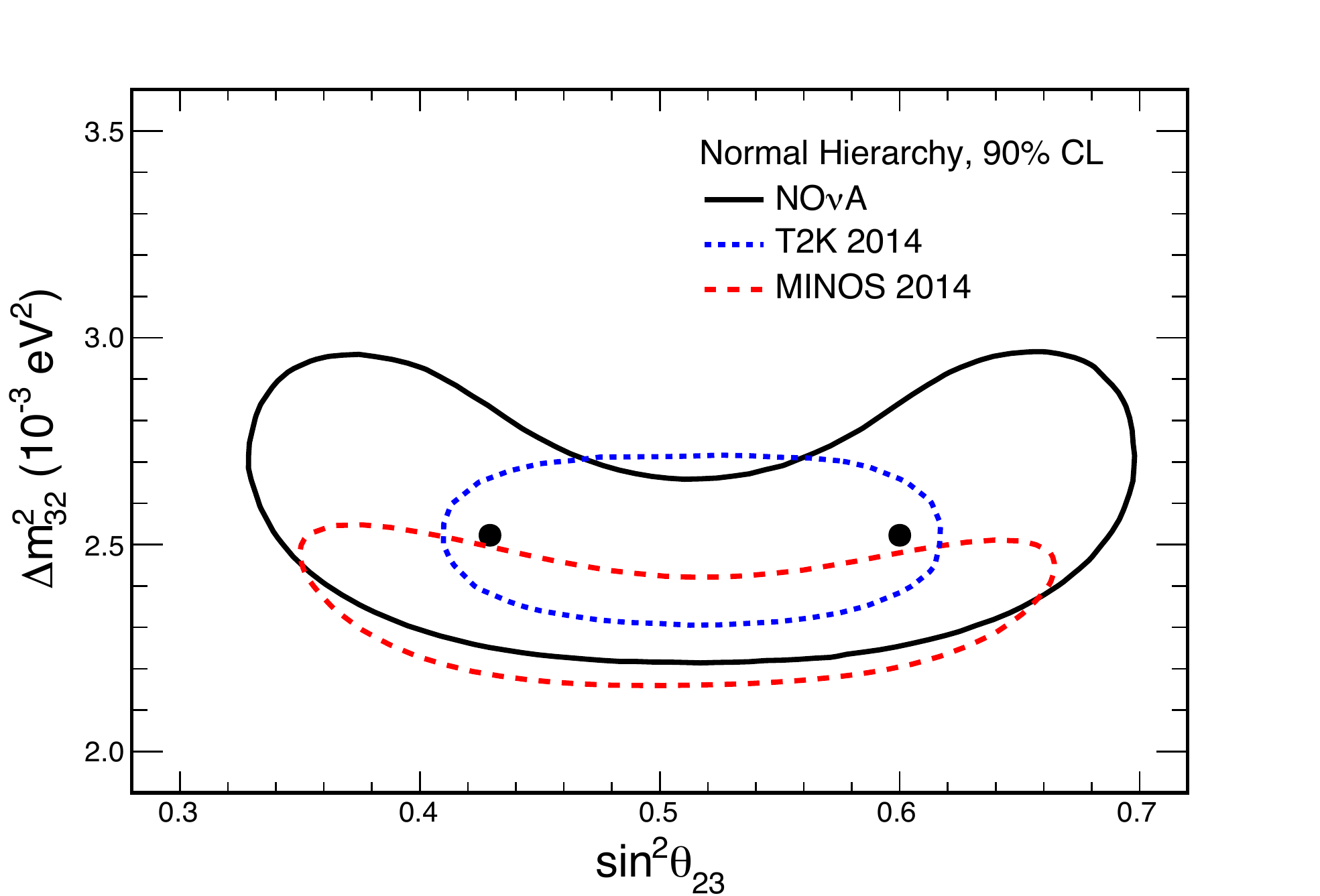}
\caption{The best-fit (solid black circles) and allowed values (solid black curve) of $\sin^2\theta_{23}$ and $\Delta m^{2}_{32}$ from this analysis assuming the normal mass hierarchy.  The dashed contour lines are results from T2K~\cite{ref:T2KCombined} and MINOS~\cite{ref:MinosCombined}.}
\label{fig:contour}
\end{figure}

Using a three-flavor neutrino oscillation model that includes matter effects, the data are fit for 
$\sin^2\theta_{23}$ and $\Delta m^2_{32}$ assuming either the normal or inverted mass hierarchy.  The fit is a %Poisson 
$\log$-likelihood maximization comparing the neutrino energy spectrum of the data to that of the extrapolated simulation over 18 bins from 0.5 to 5.0\,GeV.  Systematic effects and constraints on all other oscillation parameters are taken into account in the fit with penalty terms.  Central values and uncertainties for $\theta_{12}$ and $\Delta m^{2}_{21}$ are taken from Ref.~\cite{ref:PDG}.  We constrain $\sin^2(2\theta_{13})$ to $0.086\,\mathord{\pm}\,0.005$, a weighted average of recent results~\cite{ref:DayaBay,ref:RENOLatest,ref:DoubleChooz}.  $\delta_{CP}$ is unconstrained. The resulting allowed region, calculated using the Feldman-Cousins technique~\cite{ref:FeldmanCousins}, is shown in Fig.~\ref{fig:contour}.  1-D 68\% confidence level (CL) ranges for each of $\Delta m^2_{32}$ and $\sin^2\theta_{23}$ are obtained by maximizing the profile likelihood ratio of each parameter \cite{ref:PDG}. 

Assuming the normal hierarchy, we measure \mbox{$\Delta m^2_{32}=(2.52\,^{+0.20}_{-0.18})\mathord{\times}10^{-3}$\,eV$^{2}$}  and $\sin^2\theta_{23}$ in the 68\% CL range \mbox{$[0.38, 0.65]$}, with two statistically degenerate best-fit values of $\sin^2\theta_{23}$ of 0.43 and 0.60.
Assuming the inverted hierarchy, we measure \mbox{$\Delta m^2_{32}=(-2.56\,\mathord{\pm}\,0.19)\mathord{\times}10^{-3}$\,eV$^{2}$} and $\sin^2\theta_{23}$ in the 68\% CL range \mbox{$[0.37, 0.64]$}, with two statistically degenerate best-fit values of $\sin^2\theta_{23}$ of 0.44 and 0.59.  The best-fit parameters in both hierarchies yield a prediction of 35.4\,events in the FD.

In conclusion, the first \nova measurement of $\sin^2\theta_{23}$ and $\Delta m^{2}_{32}$ through observation of the disappearance of muon neutrinos is reported.  The results, based on less than 10\% of the planned exposure of the \nova experiment, are consistent with maximal $\theta_{23}$ mixing as well as with current results from \cite{ref:DayaBay,ref:IceCube,ref:T2KCombined,ref:MinosCombined,ref:SuperKAtmos}.

This work was supported by the US Department of Energy; the US National Science Foundation; the Department of Science and Technology, India; the European Research Council; the MSMT CR, Czech Republic; the RAS, RMES, and RFBR, Russia; CNPq and FAPEG, Brazil; and the State and University of Minnesota. We are grateful for the contributions of the staffs of the University of Minnesota module assembly facility and NOvA FD Laboratory, Argonne National Laboratory, and Fermilab. Fermilab is operated by Fermi Research Alliance, LLC under Contract No. De-AC02-07CH11359 with the US DOE.

\end{document}